\documentclass[preprin,showpacs,preprintnumbers,eqsecnum,amsmath,amssymb,twoside]{revtex4}
\usepackage{graphics,epsfig}
\usepackage{graphicx}
\usepackage{dcolumn}
\usepackage{bm}

\begin{document}
\baselineskip=0.5 cm
\title{{\bf Strong gravitational lensing in a black-hole spacetime dominated by dark energy}}

\author{Chikun Ding }
\email{dingchikun@163.com} \affiliation{Department of Physics and Information Engineering,  \\
Hunan University of Humanities,  Science and Technology,  Loudi, Hunan
417000, P. R. China}

\author{Rong-Gen Cai}
\email{cairg@itp.ac.cn}\affiliation{State Key Laboratory of Theoretical Physics,
Institute of Theoretical Physics, Chinese Academy of Sciences,
Beijing 100190, P. R. China}

\author{Changqing Liu }
\affiliation{Department of Physics and Information Engineering, \\
Hunan University of Humanities, Science and Technology,  Loudi, Hunan
417000, P. R. China}
\author{Yuanyuan Xiao}
\affiliation{Department of Physics and Information Engineering, \\
Hunan University of Humanities, Science and Technology,  Loudi, Hunan
417000, P. R. China}
\author{Liqun Jiang}
\affiliation{Department of Physics and Information Engineering, \\
Hunan University of Humanities, Science and Technology,  Loudi, Hunan
417000, P. R. China}

\vspace*{0.2cm}
\begin{abstract}
\baselineskip=0.5 cm
\begin{center}
{\bf Abstract}
\end{center}

We study the influence of phantom fields on strong field gravitational lensing. Supposing that the
gravitational field of the supermassive central object of the Galaxy is described by a phantom black hole metric, we estimate the numerical values of the coefficients and observations and find that the influence of the phantom fields is somewhat similar to that of the electric charge in a Reissner-Norstr\"{o}m black hole, i.e., the deflect angle and angular separation increase with the phantom constant $b$. However, other observations are contrary to the Reissner-Norstr\"{o}m case and show the effects of dark energy, such as (i)  compressing the usual black hole and more powerfully attracting photons, (ii) making the relativistic Einstein ring larger than that of the usual black hole, and (iii) not weakening the usual  relative magnitudes, which will facilitate observations.
\end{abstract}

 \maketitle

\vspace*{0.2cm}

\section{Introduction}

Modern observational programs---including type Ia supernovae, cosmic microwave background anisotropy, and mass power spectrum observations \cite{bachall}---indicate that the Universe is
expanding with an acceleration that is dominated to about 70\% by a peculiar kind of matter called dark energy ( characterized by negative values of the pressure-to-density ratio $\omega$), while the remaining 30\% consists of baryonic and nonbaryonic visible and dark matter. The simplest way to describe this dark energy is through the use of quintessence ($-1<\omega<-1/3$) or a phantom scalar field ($\omega\leq-1$) instead of a canonical one, that is, a scalar with a negative sign for the kinetic term in the Lagrangian \cite{bronnikov2}.  The values $\omega<-1$ should be noted because they seem to be not only admissible but even preferable for describing an increasing acceleration, which follows from the most recent estimates: $\omega=-1.10\pm0.14 (1\sigma)$ \cite{komatsu} (according to the 7-year WMAP data) and $\omega=-1.069^{+0.091}_{-0.092}$ \cite{sullivan} (mainly from data on type Ia supernovae from the SNLS3 sample). Thus, $\omega=-1$ is commonly  admitted by observations as a possible dark energy model. Through this connection, cosmological models with phantom scalar fields have gained considerable attention in recent years \cite{gannouji}.

If such a phantom scalar is part of the real field content of our Universe, it is natural to seek
its manifestations not only in cosmology but in local phenomena as well, in particular in black hole physics such as dark energy accretion onto black holes \cite{babichev},  black hole interactions with a phantom shell \cite{berezin}, the existence of regular black holes from a system of gravity coupled to these phantom fields \cite{bronnikov}, etc.
How do we test these phantom fields? The best approach would be gravitational lensing, as its resolution ratio is many orders of magnitude higher than any artificial telescope \cite{mcbreen}. Gravitational lenses are now used to determine the Hubble constant \cite{dyer}, probe the structure of a galaxies \cite{wu}, measure the density of  cosmic strings \cite{gott}, and restrict the density factor of the Universe \cite{wu}. Microlensing---such as that arising from stars and black holes---are used to probe dark matter and dark energy in the Galactic halo \cite{chang}, etc.,
so we can use it here to probe the existence and distribution of  dark energy via the influence of a phantom scalar on the gravitational field, i.e., on a black hole lens' behavior.

The earlier studies of gravitational lensing were developed in
the weak-field approximation \cite{Schneider}---\cite{RDB}. It is
enough for us to investigate the properties of gravitational lensing
by ordinary stars and galaxies. However, when the lens is a black
hole, a strong-field treatment of gravitational lensing
\cite{Darwin,Vir,Vir1,Vir2,Vir3,Fritt} is needed instead. Virbhadra and
Ellis \cite{Vir1} found that near the line connecting the source and
the lens, an observer would detect two infinite sets of faint
relativistic images on each side of the black hole. These relativistic images could
provide a profound verification of alternative theories of gravity.
Thus, the study of strong gravitational lensing has become
appealing in recent years. On the basis of the Virbhadra-Ellis lens equation \cite{Vir2,Vir3},
Bozza \cite{Bozza2} extended the analytical method of lensing for a
general class of static and spherically symmetric spacetimes and
showed that the logarithmic divergence of the deflection angle at
the photon sphere is a common feature. Bhadra
\textit{et al.} \cite{Bhad1}\cite{Sarkar} considered the
Gibbons-Maeda-Garfinkle-Horowitz-Strominger black hole lensing. Eiroa \textit{et al} \cite{Eirc1} studied the
Reissner-Nordstr\"{o}m black hole lensing. Konoplya \cite{Konoplya1}
 studied the corrections to the deflection angle and time delay
of black hole lensing immersed in a uniform magnetic
field. Majumdar \cite{Muk} investigated the dilaton-de Sitter black hole lensing. Perlick
\cite{Per} obtained an exact lens equation and used it to study
 Barriola-Vilenkin monopole black hole lensing. Virbhadra {\it et al.} studied the relativistic images of spherically symmetric black hole lensing without any approximations (i.e., the strong- or weak-field treatments) \cite{virbhadra}. S. Chen studied Kehagias-Sfetsos black hole lensing \cite{chen}.  Bin-Nun \cite{bin} studied the strong gravitational lensing by Sgr A*, G. N. Gyulchev studied phantom black hole lensing \cite{gyulchev}, and
so on.

This paper is organized as follows. In Sec. II we briefly  review the regular phantom black holes. In Sec. III we adopt
Bozza's method and obtain the deflection angles for light rays
propagating in the phantom black hole
spacetime. In Sec. IV, we discuss the time delay of light seen from images. In Sec. V we suppose that the gravitational field of the
supermassive black hole at the center of our Galaxy can be described
by this metric and then obtain the numerical results for the
observational gravitational lensing parameters defined in Secs. III and IV.
Then, we make a comparison between the properties of gravitational
lensing in the phantom black hole and
Reissner-Norstr\"{o}m metrics. In Sec. VI we present a summary.

\section{Phantom black holes}
\begin{figure}[ht]
\begin{center}
\includegraphics[width=7.0cm]{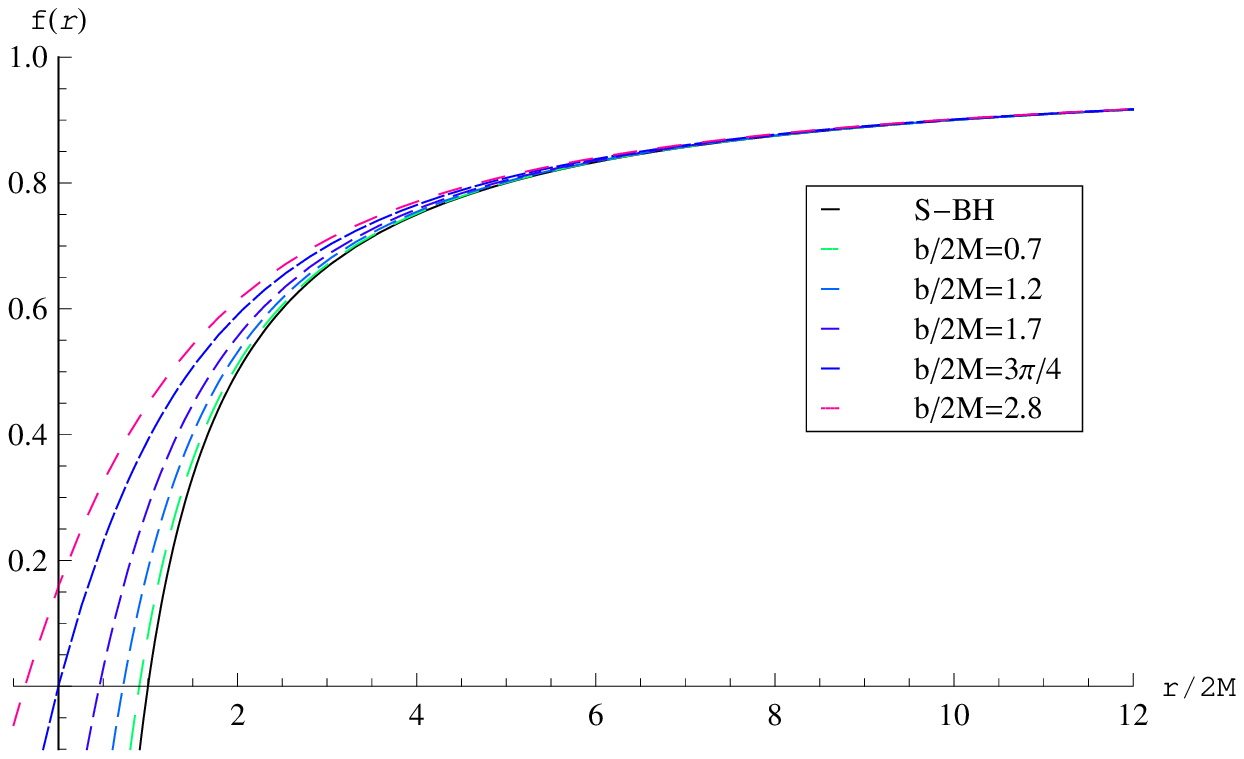}\;\;\;\;\;\includegraphics[width=7.0cm]{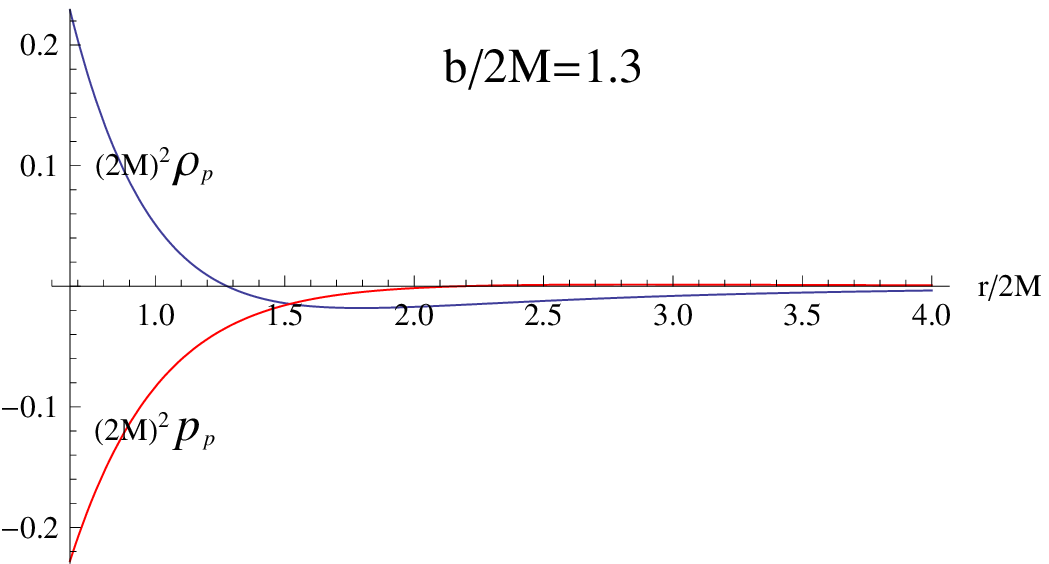}
\caption{Left: The metric functions of the
regular phantom black hole for different values of
$b$.  The solid line is described by the expression
$f(r)=1-2M/r$ for a Schwarzschild black hole, ``S-BH." Right: The energy density $\rho_p$ and pressure $p_p$ for a phantom field with $b/2M=1.3$.}\label{fp1}
 \end{center}
 \end{figure}

Consider the Lagrangian
\begin{equation} L=\sqrt{-g}\Big[-\frac{R}{8\pi G}+\epsilon g^{\alpha\beta}\phi_{;\alpha}\phi_{;\beta}-2V(\phi)\Big],
\end{equation}
which includes a scalar field, in general, with some potential $V(\phi)$; $\epsilon$
distinguishes normal, canonical scalar fields ($\epsilon= +1$) and
phantom fields ($\epsilon =-1$).
The static, spherically symmetric metric for phantom scalar fields can
be written in the form \cite{bronnikov}
\begin{eqnarray}
ds^2 = -f(r)\,dt^2 + \frac{dr^2}{f(r)} + (r^2 +b^2)(d\theta^2
+\sin^2\theta d\phi^2)\,, \label{metric}
\end{eqnarray}
with
\begin{eqnarray}
\label{sol1} f(r)=1- \frac{3M}{b}\, \Big[\big(\frac{\pi}{2}-\text{arctan}\frac{r}{b}\big)
\big(1+\frac{r^2}{b^2}\big)-\frac{r}{b}\Big],
\end{eqnarray}
where $M$ is the black hole's mass defined in the usual way, $b$ is a positive constant relative to the  charge of phantom scalar fields (termed the phantom constant), and its potential is
\begin{equation}
\frac{\phi}{\sqrt{2}}\equiv\psi= \,\text{arctan}\frac{r}{b},\,\, V=\frac{3 M}{b^3}\Big[(\frac{\pi}{2}-\psi)(3-2\, \text{cos}^2\psi)-3\, \text{sin}\psi\text{cos}\psi\Big].
\end{equation}
This metric behavior is controlled by two integration constants: $b,$ and $M$. When $M=0$, this is an Ellis wormhole. If $M<0$, it is a wormhole which is asymptotically flat at $r\rightarrow\infty$ and which has an anti-de Sitter metric at $r\rightarrow-\infty$. When $M>0$, it is a regular black hole whose curvature scalar at the origin is
 \begin{equation}\label{ricci}
R_{\mu\nu\tau\rho}R^{\mu\nu\tau\rho}=\frac{3(4b^2-8bM\pi+9\pi^2M^2)}{b^6},
\end{equation}
  and it has a Schwarzschild-like causal structure at large $r$. In Fig. \ref{fp1},  we show the behavior of the black holes' metric functions (\ref{sol1}) and the energy density, and pressure for the phantom field. In Table \ref{tab1} its horizon, pressure, and pressure-to-density ratio  at horizon for different values of $b$ are listed.
 \begin{table}
 \caption{Numerical values for the radius of the single event horizon of a phantom black hole, the pressure-to-density ratio $\omega_p$, and the pressure $p_+$ for phantom fields near the horizon for different values of the phantom constant $b$. Here,  $\omega_p=-1+\omega\times10^{-7}$.}\label{tab1}
 \begin{center}
\begin{tabular}{cccccccccccc}
\hline \hline $b/2M$ &0.02&0.1 &0.4&0.7 &1.0 &1.3&1.6&1.9&2.2&2.3&$3\pi/4 $\\
\hline
 $r_+/2M$&0.99992&0.998001& 0.968143&0.90329&0.805076&0.675537&0.51679&0.330846
 &0.119525&0.043730&0 \\
 $\omega $&3.2612&$-7.1203$&4.0101&3.4691&1.02028&$-$1.71816&1.05262&0.996388
 &1.17354&$-$0.03441&0.0
 \\$-(2M)^2p_+$&0.000160&0.003966&0.056039&0.133647&0.194823&0.224111&0.226573&0.212936
 &0.191999&0.184419&0.180127\\
\hline\hline
\end{tabular}
\end{center}
\end{table}

From Table \ref{tab1} we can see that if the phantom constant is small or even if $b\rightarrow0$, then the black hole behaves as a Schwarzschild black hole [it cannot recover a Schwarzschild black hole due to the fact that $b\neq0$ from Eq. (\ref{ricci}) ]. In this case we can call it a phantom Schwarzschild black hole. When $b$ increases, the radius of the horizon decreases and $-p_+$ increases, which indicates a stronger effect from dark energy.  The pressure-to-energy density ratio $\omega_p$ of this dark energy is around $-1$, which is coincident with present observations \cite{komatsu,sullivan}. The expressions for $\omega_p,\;\rho_p,$ and $p_p$ are included in the Appendix. Table \ref{tab1} also shows that phantom fields affect the size of the black hole.  In addition, the phantom constant $b$ behaviors somewhat like the electric charge $q$ in a Reissner-Norstr\"{o}m black hole (whose external horizon  decreases with $q$), so we can compare phantom black hole lensing to Reissner-Norstr\"{o}m lensing.
The line element (\ref{metric}) describes the geometry of a
phantom black hole and should give us useful insights about
possible dark energy effects on strong gravitational
lensing.
\section{Deflection angle in the phantom black hole spacetime}
From this section and hereafter, we set $b/2M=b,\;r/2M=r,\;u/2M=u,\;q/2M=q$, and rewrite the metric
(\ref{metric}) as
\begin{eqnarray}
ds^2=-A(r)dt^2+B(r)dr^2+C(r)\Big(d\theta^2+\sin^2\theta
d\phi^2\Big),\label{grm}
\end{eqnarray}
with
\begin{eqnarray}
A(r)=f(r), \;\;\;\;B(r)&=&1/f(r),\;\;\;\; C(r)=r^2 +b^2.
\end{eqnarray}
The deflection angle for the photon coming from infinity can be
expressed as
\begin{eqnarray}
\alpha(r_0)=I(r_0)-\pi,
\end{eqnarray}
where $r_0$ is the closest approach distance and $I(r_0)$ is
\cite{Vir2,Vir3}
\begin{eqnarray}
I(r_0)=2\int^{\infty}_{r_0}\frac{\sqrt{B(r)}dr}{\sqrt{C(r)}
\sqrt{\frac{C(r)A(r_0)}{C(r_0)A(r)}-1}}.\label{int1}
\end{eqnarray}
It is easy to see that as the parameter $r_0$ decreases the deflection
angle increases. At a certain a point, the deflection angle will become
$2\pi$, which means that the light ray will make a complete loop around
the compact object before reaching the observer. When $r_0$ is equal
to the radius of the photon sphere, the deflection angle diverges
and the photon is captured.

The photon-sphere equation is given by \cite{Vir2,Vir3}
\begin{eqnarray}
\frac{C'(r)}{C(r)}=\frac{A'(r)}{A(r)},\label{root}
\end{eqnarray}
which admits at least one positive solution, and then the largest
real root of Eq.(\ref{root}) is defined  as the radius of the photon
sphere. Using the phantom black hole metric
(\ref{metric}),  Eq. (\ref{root}) is
\begin{eqnarray}
\frac{2r}{r^2+b^2}=\frac{\frac{3}{b^2}\big[1-\big(\frac{\pi}{2}-\arctan\frac{r}{b}\big)\frac{r}{b}
\big]}{1- \frac{3}{2b}\, \big[\big(\frac{\pi}{2}-\arctan\frac{r}{b}\big)
\big(1+\frac{r^2}{b^2}\big)-\frac{r}{b}\big]}.
\end{eqnarray}
After a simple calculation, this can be simplified to $2r=3$, so that
the radius of the photon sphere can be given
 by
\begin{eqnarray}\label{rpseq}
r_{ps}=\frac{3}{2},
\end{eqnarray}
which is the same as that of a Schwarzschild black hole and is independent of the constant $b$. It tells us that (i) no matter how the phantom fields are distributed the photon sphere stays the same, and (ii) we cannot distinguish merely from the photon sphere whether dark energy exist or not.

Following the method developed by Bozza\footnotemark\footnotetext{Though Bozza's prescriptions have been subjected to much criticism of inaccuracy \cite{virbhadra}, it can give us a clear picture of strong gravitational lensing from an analytic point of view. For numerical works without any approximations  on spherically symmetric black hole lensing please see Ref. \cite{virbhadra}.} \cite{Bozza2,chen},we
define a variable
\begin{eqnarray}
z=1-\frac{r_0}{r},
\end{eqnarray}
and obtain
\begin{eqnarray}
I(r_0)=\int^{1}_{0}R(z,r_0)f(z,r_0)dz,\label{in1}
\end{eqnarray}
where
\begin{eqnarray}
R(z,r_0)&=&\frac{2r_0\sqrt{A(r)B(r)C(r_0)}}{C(r)(1-z)^2}
=\frac{2r_0\sqrt{r_0^2+b^2}}{(r^2+b^2)(1-z)^2},
\end{eqnarray}
\begin{eqnarray}
f(z,r_0)&=&\frac{1}{\sqrt{A(r_0)-A(r)C(r_0)/C(r)}}.
\end{eqnarray}
The function $R(z, r_0)$ is regular for all values of $z$ and $r_0$.
However, $f(z, r_0)$ diverges as $z$ tends to zero. Thus, we split
the integral (\ref{in1}) into two parts
\begin{eqnarray}
I_D(r_0)&=&\int^{1}_{0}R(0,r_{ps})f_0(z,r_0)dz, \nonumber\\
I_R(r_0)&=&\int^{1}_{0}[R(z,r_0)f(z,r_0)-R(0,r_{ps})f_0(z,r_0)]dz
\label{intbr},
\end{eqnarray}
where $I_D(r_0)$ and $I_R(r_0)$ denote the divergent and regular
parts in the integral (\ref{in1}), respectively. To find the order
of divergence of the integrand, we expand the argument of the square
root in $f(z,r_0)$ to the second order in $z$ and obtain the
function $f_0(z,r_0)$,
\begin{eqnarray}
f_0(z,r_0)=\frac{1}{\sqrt{p(r_0)z+q(r_0)z^2}},
\end{eqnarray}
where
\begin{eqnarray}
p(r_0)=\frac{r_0(2r_0-3)}{r_0^2+b^2},\;\;
q(r_0)=\frac{r_0^2}{r_0^2+b^2}+\frac{r_0(2r_0-3)(b^2-r_0^2)}{(r_0^2+b^2)^2}.
\end{eqnarray}
When $r_0$ is equal to the radius of the photon sphere $r_{\text{ps}}$, the
coefficient $p(r_0)$ vanishes, and the leading term of the divergence
in $f_0(z,r_0)$ is $z^{-1}$; thus, the integral (\ref{in1}) diverges
logarithmically. Close to the divergence, Bozza \cite{Bozza2} found
that the deflection angle can be expanded in the form
\begin{eqnarray}
\alpha(\theta)=-\bar{a}\log{\bigg(\frac{\theta
D_{\text{OL}}}{u_{\text{ps}}}-1\bigg)}+\bar{b}+O(u-u_{\text{ps}}),
\end{eqnarray}
where
\begin{eqnarray}
&\bar{a}&=1, \nonumber\\
&\bar{b}&=
-\pi+b_R+\bar{a}\log{\frac{4q^2(r_{\text{ps}})\big[2A(r_{\text{ps}})-(r_{\text{ps}}
^2+b^2)A''(r_{\text{ps}})\big]}{
p^{'2}(r_{\text{ps}})u_{\text{ps}}\sqrt{A^3(r_{ps})(r_{\text{ps}}^2+b^2)}}}, \nonumber\\
&b_R&=I_R(r_{\text{ps}}),\;\;\;\;\;p'(r_{ps})=\frac{dp}{dr_0}\big|_{r_0=r_{\text{ps}}},\;\;\;\;\;u_{ps}
=\frac{\sqrt{r_{\text{ps}}^2+b^2}}{\sqrt{A(r_{\text{ps}})}}.
\end{eqnarray}
$D_{\text{OL}}$ denotes the distance between the observer and the
gravitational lens, and $\bar{a}$ and $\bar{b}$ are the so-called the strong-field limit coefficients which depend on the metric functions
evaluated at $r_{\text{ps}}$. In general, the coefficient $b_R$ cannot be
calculated analytically, but in this case it can be evaluated
numerically.

\begin{figure}[ht]
\begin{center}
 \includegraphics[width=7cm]{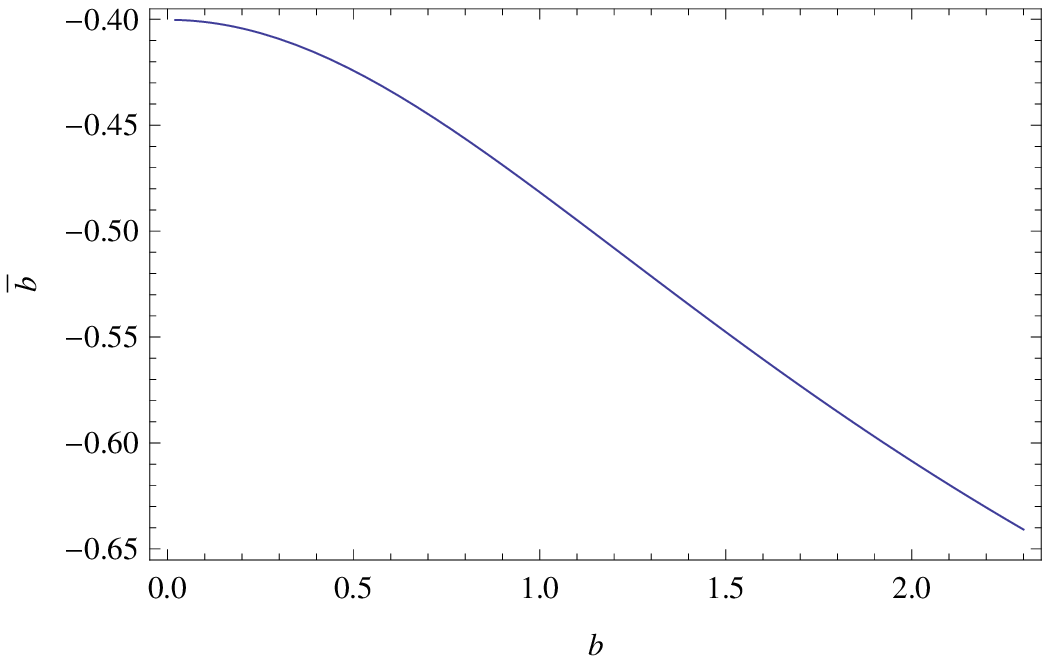}\;\;\;\;\includegraphics[width=7cm]{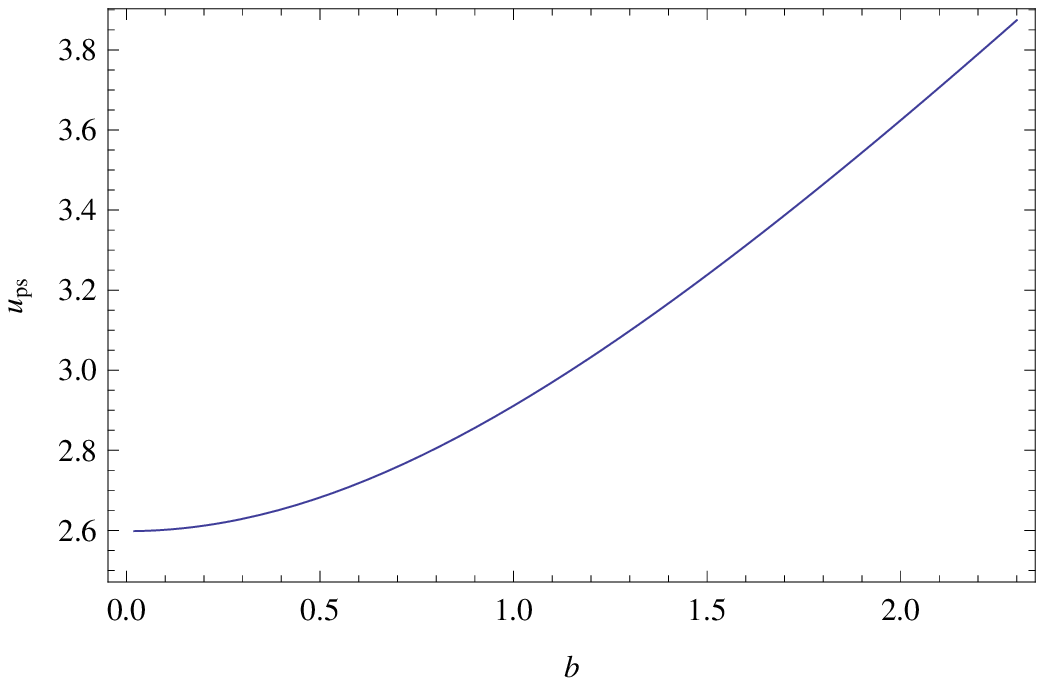}
\\ \vspace{0.6cm}
 \includegraphics[width=7cm]{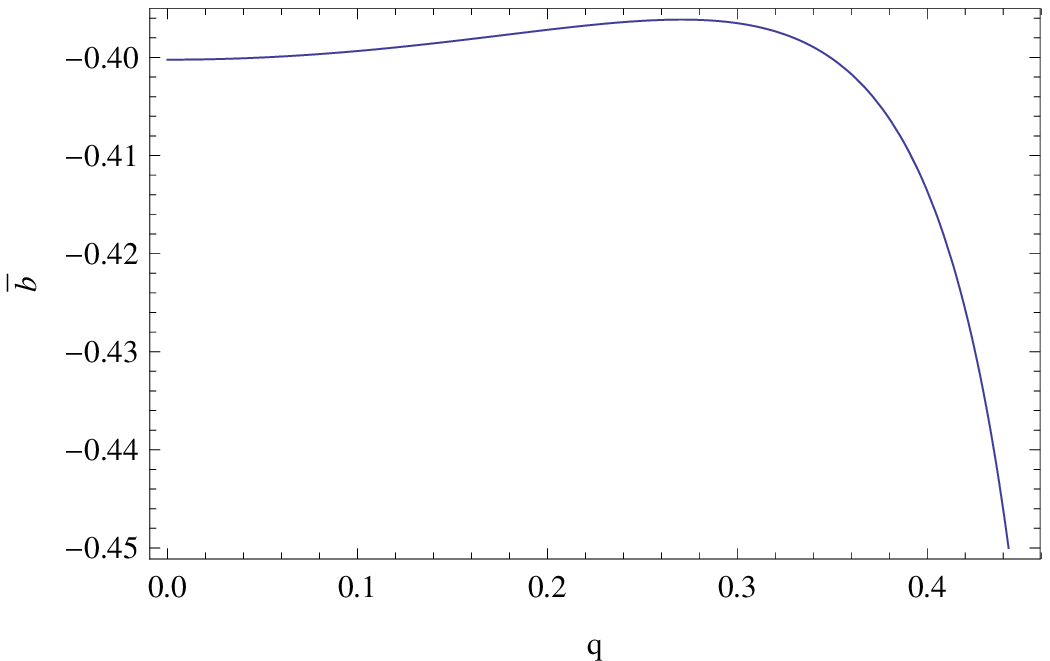}\;\;\;\;\includegraphics[width=7cm]{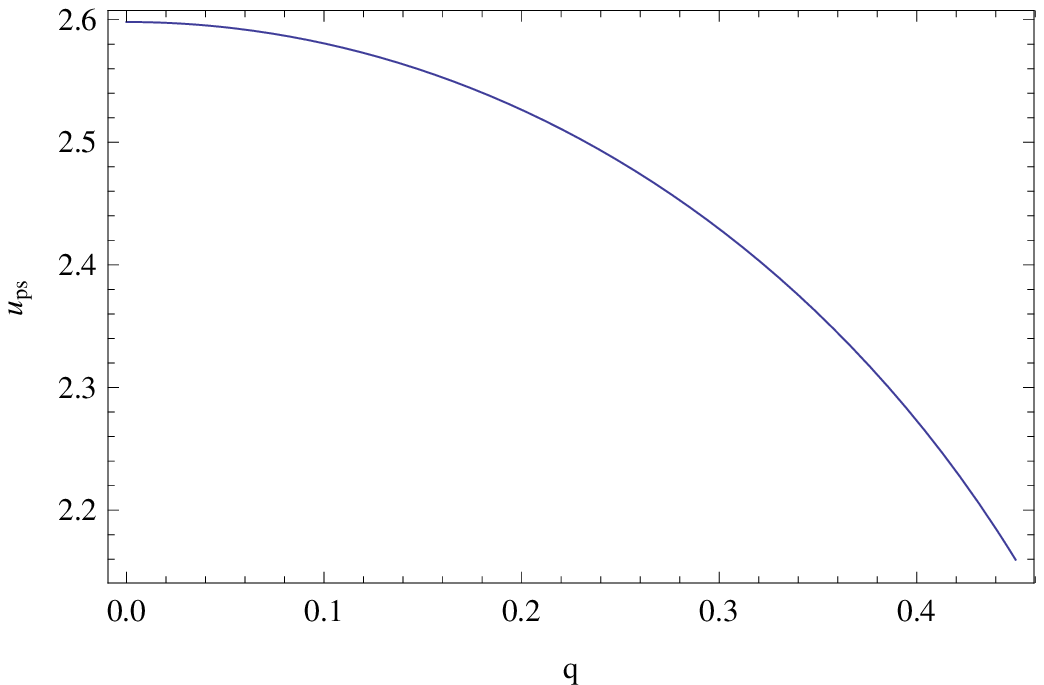}
\caption{The coefficient of the strong-field limit
$\bar{b}$ and the minimum impact parameter $u_{\text{ps}}$ vs
the phantom constant $b$ in the
phantom black hole spacetime (upper panels)
and vs $q$ in the Reissner-Nordstr\"{o}m black hole spacetime (lower panels). The values of the coefficient of
Reissner-Nordstr\"{o}m lensing come from Ref. \cite{Bozza2}.}\label{fp2}
 \end{center}
 \end{figure}

\begin{figure}[ht]
\begin{center}
 \includegraphics[width=7cm]{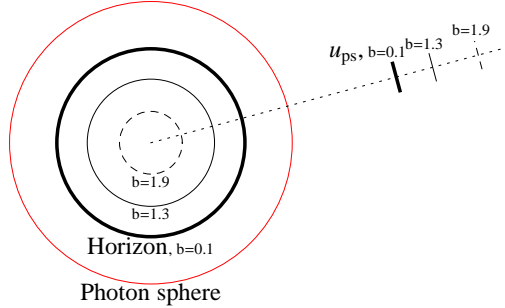}
\caption{The position of the photon sphere, black hole horizon, and minimum impact parameter $u_{\text{ps}}$ for different values of the phantom constant $b$.}\label{fp5}
 \end{center}
 \end{figure}

Then we can now obtain the $\bar{b}$ and $u_{\text{ps}}$, which we show
in Fig. \ref{fp2}.
 Figure \ref{fp2} show us that as
$b$ increases  the coefficient $\bar{b}$ always decreases, whereas in the Reissner-Norstr\"{o}m case there is a region of increase with the electric charge $q$. Also, the minimum impact parameter $u_{\text{ps}}$
increases, which is contrary to the case in the Reissner-Norstr\"{o}m
black hole spacetime. This behavior will greatly affect both the deflect angle and the angular separation.
Figure \ref{fp5} shows the relative position of the photon sphere, black hole horizon and minimum impact parameter with different $b$. Bigger $u_{ps}$ indicates that, in more farther place, the deflect angle of photons will also diverge.  It is easy to see that a larger phantom constant corresponds to a stronger interaction of dark energy on the spacetime, which causes it more curved, i.e., compresses the black hole and more powerfully attracts photons. In principle we can distinguish a phantom black hole from the
Reissner-Nordstr\"{o}m black hole and probe the value of the phantom constant by using strong field gravitational lensing.

\begin{figure}[ht]
\begin{center}
\includegraphics[width=7.0cm]{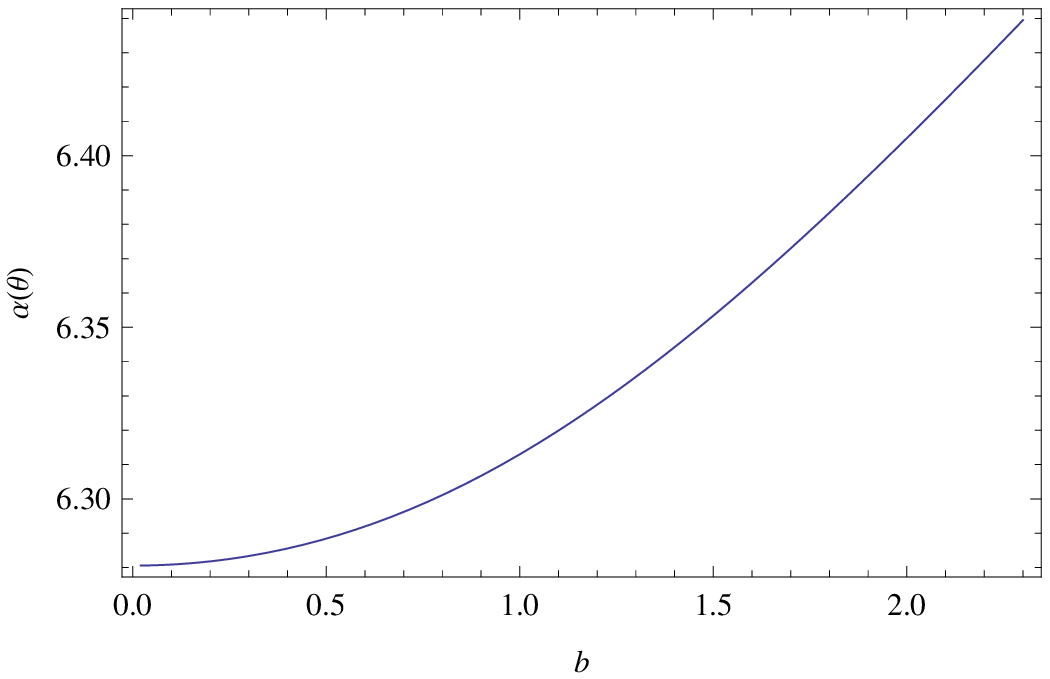}\;\;\;\;
 \includegraphics[width=7.0cm]{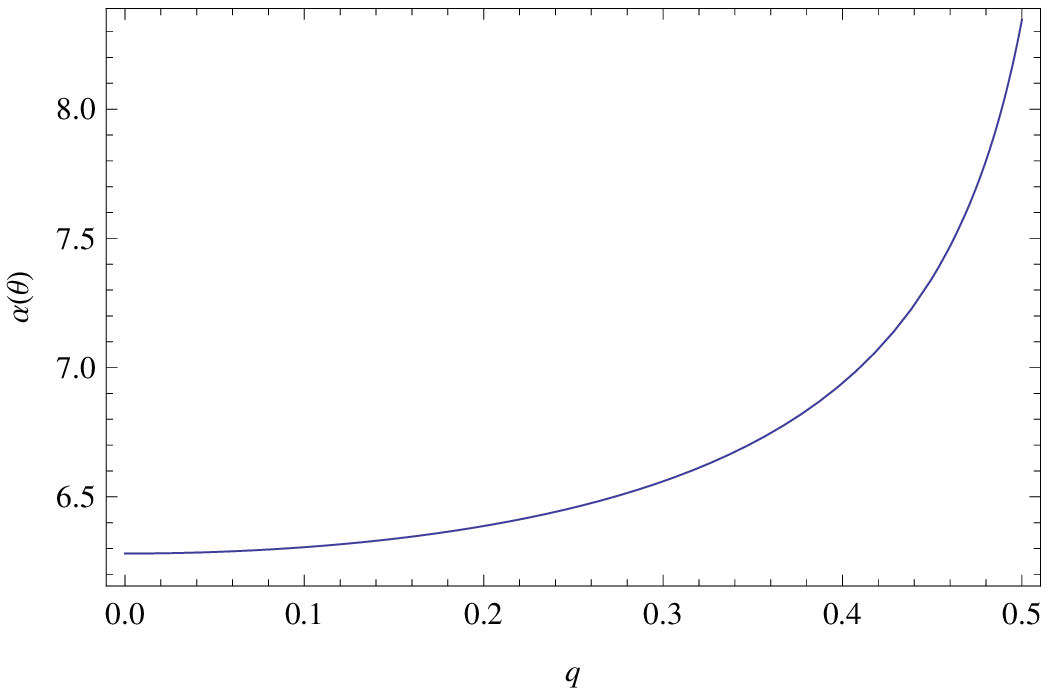}
\caption{ Deflection angles in the phantom
black hole (left) and Reissner-Nordstr\"{o}m (right) spacetime evaluated at $u=u_{\text{ps}}+0.00326$. }\label{fp3}
 \end{center}
 \end{figure}

Figure \ref{fp3} shows the deflection angle $\alpha (\theta)$ evaluated at
$u=u_{\text{ps}}+0.00326$. It indicates that the presence of $b$
increases the deflection angle $\alpha (\theta)$ for the light
propagated in the phantom black hole spacetime, which is similar to the electric charge $q$ in the Reissner-Nordstr\"{o}m case.
Comparing with the Reissner-Nordstr\"{o}m case, we could extract the
information about the size of the phantom constant
$b$ by using strong field gravitational lensing.

Assuming that the source, lens, and observer are highly aligned, the
lens equation in strong gravitational lensing can be written as
\cite{Bozza1}
\begin{eqnarray}\label{lenseq}
\tan\beta=\tan\theta-\frac{D_{\text{LS}}}{D_{\text{OS}}}\big[\tan\theta+\tan(\Delta\alpha_{n}-\theta)\big],
\end{eqnarray}
where $D_{\text{LS}}$ is the distance between the lens and the source,
$D_{\text{OS}}=D_{\text{LS}}+D_{\text{OL}}$, $\beta$ is the angular separation between
the source and the lens, $\theta$ is the angular separation between
the image and the lens, and $\Delta\alpha_{n}=\alpha-2n\pi$ is the
offset of deflection angle with $n$ an integer. The position of
the $n$th relativistic image can be approximated as
\begin{eqnarray}\label{thetan}
\theta_n=\theta^0_n+\frac{u_{\text{ps}}e_n(\beta-\theta^0_n)D_{\text{OS}}}{\bar{a}D_{\text{LS
}}D_{\text{OL}}},
\end{eqnarray}
where
\begin{eqnarray}
e_n=e^{\frac{\bar{b}-2n\pi}{\bar{a}}},
\end{eqnarray}
$\theta^0_n$ are the image positions corresponding to
$\alpha=2n\pi$.  The magnification of the $n$th relativistic image is
given by
\begin{eqnarray}
\mu_n=\frac{u^2_{\text{ps}}e_n(1+e_n)D_{\text{OS}}}{\bar{a}\beta D_{\text{LS}}D^2_{\text{OL}}}.
\end{eqnarray}
If $\theta_{\infty}$ represents the asymptotic position of a set of
images in the limit $n\rightarrow \infty$, the minimum impact
parameter $u_{\text{ps}}$ can be simply obtained as
\begin{eqnarray}
u_{\text{ps}}=D_{\text{OL}}\theta_{\infty}.
\end{eqnarray}
In the simplest situation, we consider only that the outermost image
$\theta_1$ is resolved as a single image and all the remaining ones
are packed together at $\theta_{\infty}$. Then the angular
separation between the first image and the other ones can be expressed
as
\begin{eqnarray}
s=\theta_1-\theta_{\infty},
\end{eqnarray}
and the ratio of the flux from the first image and that from the
all other images is given by
\begin{eqnarray}
\mathcal{R}=\frac{\mu_1}{\sum^{\infty}_{n=2}\mu_{n}}.
\end{eqnarray}
For a highly aligned source, lens, and observer geometry, these
observables can be simplified as
\begin{eqnarray}
&s&=\theta_{\infty}e^{\frac{\bar{b}-2\pi}{\bar{a}}},\nonumber\\
&\mathcal{R}&= e^{\frac{2\pi}{\bar{a}}}.
\end{eqnarray}
The strong deflection limit coefficients $\bar{a}$, $\bar{b}$ and
the minimum impact parameter $u_{\text{ps}}$ can be obtained by
measuring $s$, $\mathcal{R}$ and $\theta_{\infty}$. Then, comparing
their values with those predicted by the theoretical models, we can
identify the nature of the black hole lens.

\section{Time delay in the phantom black hole spacetime}
In this section we consider the time delay of light seen from images.
Weinberg \cite{weinberg} obtained the time required for light to travel from a source at coordinates
($r,\;\theta=\pi/2,\;\varphi=\varphi_1$) to the closest point of approach (to the lens) at coordinates
($r_0,\;\theta=\pi/2,\;\varphi=\varphi_2$) by solving null geodesic equations
for general static spherically symmetric spacetime. A straightforward calculation for the metric (\ref{metric}) gives the time delay as \cite{virbhadra}
\begin{eqnarray}\label{time}
\tau(r_0)=2M\Big[\int^{\chi_s}_{r_0}\frac{dr}{F(r)}+\int^{\chi_o}_{r_0}\frac{dr}{F(r)}\Big]
-D_{\text{OS}}\sec\beta,\;\;F(r)=f(r)\sqrt{1-\frac{f(r)(r_0^2+b^2)}
{f(r_0)(r^2+b^2)}},
\end{eqnarray}
with
\begin{eqnarray}
\chi_s=\frac{D_{\text{OS}}}{2M}\sqrt{(D_{\text{LS}}/D_{\text{OS}})^2+\tan^2\beta},\;\;\chi_o=\frac{D_{OL}}{2M}.
\end{eqnarray}
In the next section we will use Eqs. (\ref{thetan}), (\ref{lenseq}), and (\ref{time})  to obtain the numerical values for the offset of the deflection angle $\Delta\alpha_{1p}$ and the time delay $\tau_{1p}$ of the first relativistic images (on the same side as the primary image).
\section{Numerical estimation of observational gravitational lensing parameters}

In this section---supposing that the gravitational field of the
supermassive black hole at the Galactic Center of the Milk Way can be
described by the phantom black hole metric---we
estimate the numerical values for the coefficients and observables
of strong gravitational lensing, and then we study the effect of
the phantom constant $b$ on the
gravitational lensing.

The mass of the central object of our Galaxy is estimated to be
$2.8\times 10^6M_{\odot}$ and its distance is around $8.5$ kpc \cite{rich}. For
different values of $b$, the numerical values of the minimum impact
parameter $u_{\text{ps}}$, the angular position of the asymptotic
relativistic images $\theta_{\infty}$, the angular separation $s$,
and the relative magnification of the outermost relativistic image
with the other relativistic images $r_{m}$ are listed in the Table
\ref{tab2}.
\begin{table}[h]
\caption{Numerical estimations for the main observables and the strong-field limit coefficients for a black hole at the center of our Galaxy,
which is assumed to be described by the phantom black hole metric. $R_s$ is the Schwarzschild radius and
$r_m=2.5\log{\mathcal{R}}$.}\label{tab2}
\begin{center}
\begin{tabular}{ccccccccccc}
\hline \hline  &\multicolumn{6}{c}{Phantom black hole}&\multicolumn{4}{c} {Reissner-Nordstr\"{o}m black hole} \\
 &$b=0.02$&$b=0.1$& $b=0.7$&$b=1.3$&$b=1.9$&$b=2.3$&$q=0.1$&$q=0.2$&$q=0.3$&$q=0.4$ \\ \hline
$\theta_\infty(\mu$ arc sec)&16.8708&16.8923&17.9149&20.1181&23.009&25.1555&16.7565&16.405&15.7743&14.759 \\
$s(\mu$ arc sec)&0.021112&0.021119&0.021445&0.022306&0.023651&0.024749&0.021635&0.0234359&0.027538&0.037984\\
$r_m$& 6.82188&6.82188&6.82188&6.82188&6.82188&6.82188&6.79094&6.68985&6.48575&6.07378 \\
$u_m/R_s$ &2.59821&2.60154&2.75902&3.09832&3.54355&3.87412&2.58062&2.52649&2.42935&2.27299\\
$\bar{a}$&1&1&1&1&1&1&1.00456&1.01974&1.05183&1.12317\\
$\bar{b}$&$-0.40027$&$-0.40125$&$-0.44471$&$-0.52133$&$-0.59706$&$-0.64085$
&$-0.39935$&$-0.39718$&$-0.39651$&$-0.41364$
 \\
\hline\hline
\end{tabular}
\end{center}
\end{table}\begin{figure}[ht]
\begin{center}
\includegraphics[width=7.0cm]{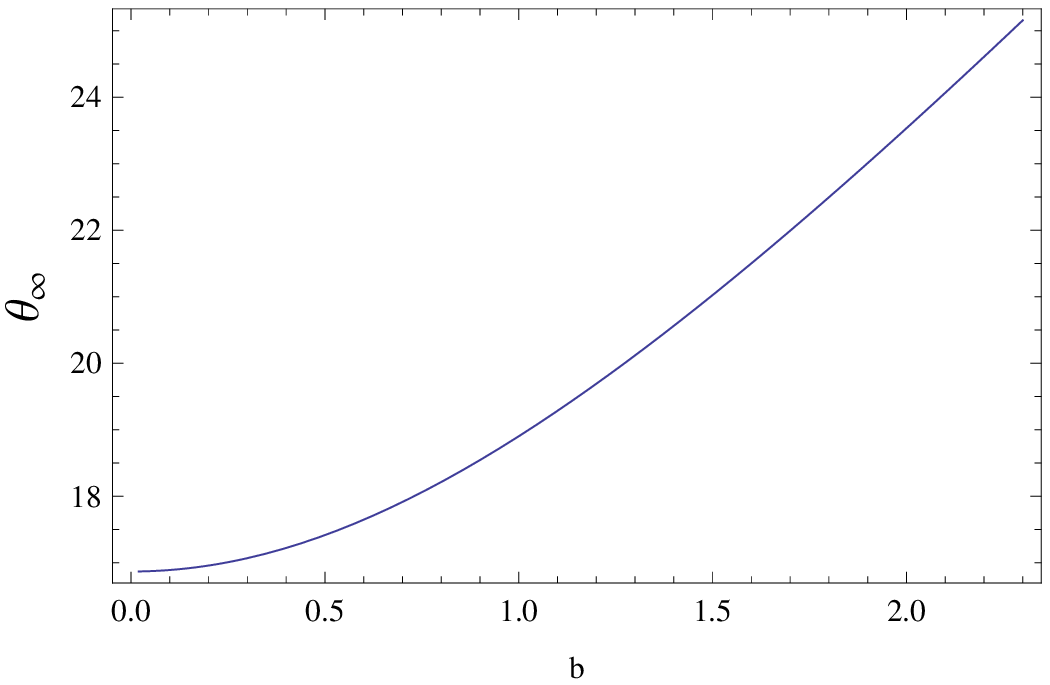}\;\;\;\;\;
 \includegraphics[width=7.0cm]{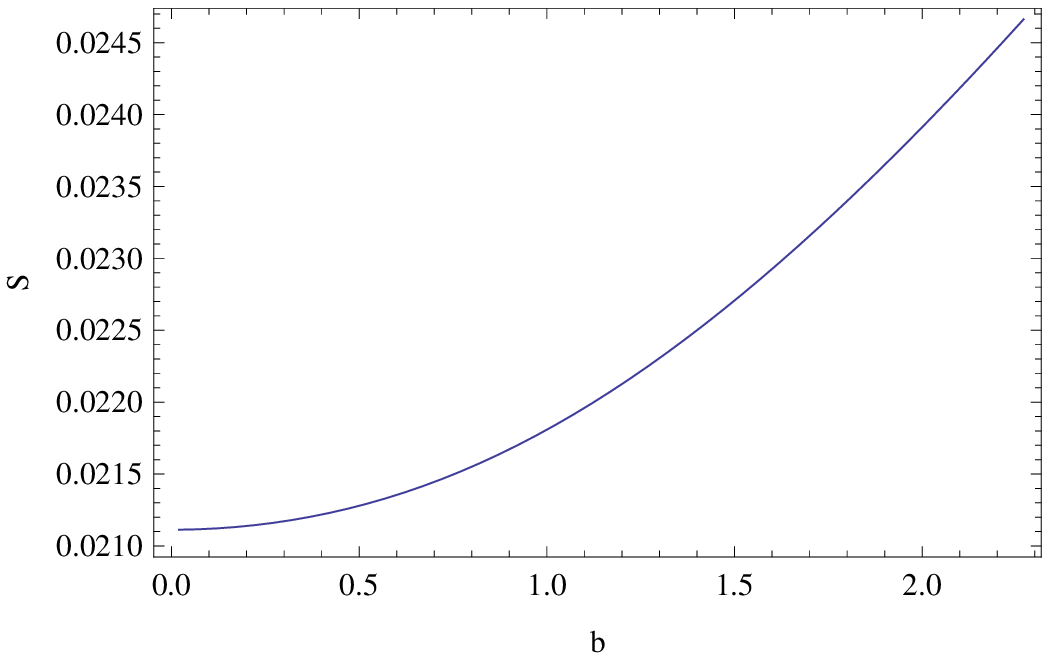}\\  \vspace {0.6cm}
 \includegraphics[width=7.0cm]{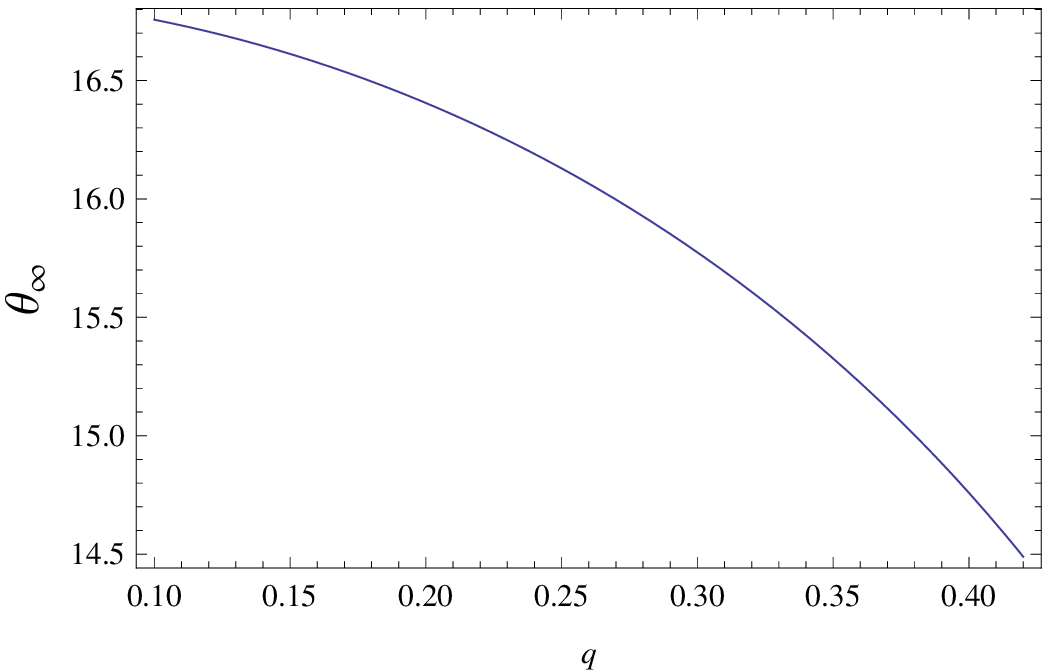}\;\;\;\;\;
 \includegraphics[width=7.0cm]{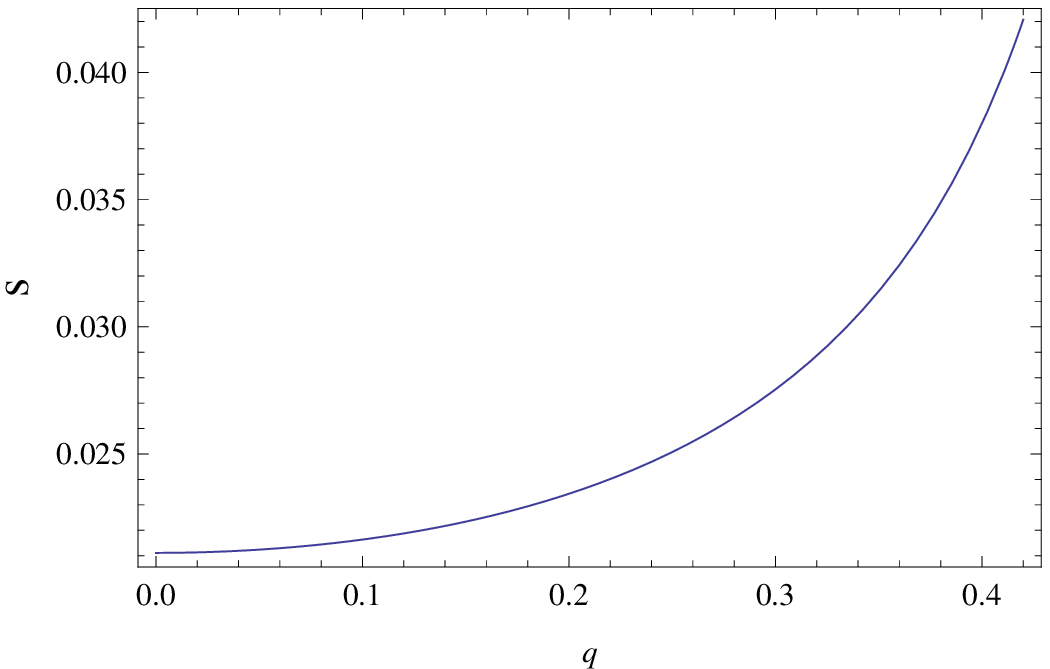}
\caption{Strong gravitational lensing by the Galactic Center black
hole. We plot the values of the angular position
$\theta_{\infty}$, the relative magnitudes $r_m$, and the angular
separation $s$ vs the phantom constant $b$ in the
phantom black hole spacetime (upper panels)
and vs $q$ in the Reissner-Norstr\"{o}m black hole (lower
panels). } \label{f3}
 \end{center}
 \end{figure}

It is easy to obtain that our results reduce to those in the
 Schwarzschild black hole spacetime as
$b\rightarrow0$. Moreover, from the Table \ref{tab2}, we also find
that as the parameter $b$ increases, the minimum impact
parameter $u_{\text{ps}}$ and the angular position of the relativistic images
$\theta_{\infty}$ increase as well, which is contrary to the Reissner-Norstr\"{o}m case. The appearance of  $\theta_{\infty}$ tells us that the relativistic Einstein ring is much bigger than teh Schwarzschild lensing. The angular separation $s$ appears to  behave similarly as in the Reissner-Norstr\"{o}m case. Also, the relative magnitude $r_m$ stays the same as in Schwarzschild lensing, that is to say, the relative flux of the first image is not affected by phantom scalar fields (dark energy).

From Figure \ref{f3}, we can see that for the phantom
black hole an increase of the parameter $b$ causes an increase of both the angular
position $\theta_{\infty}$ and the angular separation $s$. This means that the
bending angle is bigger in the phantom black hole spacetime.
 In
order to identify the nature of the lensing of these two compact objects,
it is necessary to measure the angular separation $s$ and the
relative magnification $r_m$ in the astronomical observations.
Table \ref{tab2} tells us that the resolution of the extreme angular separation image
is $\sim 0.025$ $\mu$ arcsecond, which is too small. However, as new
 technologies are developed the effects of the phantom constant $b$ on gravitational lensing may be
detected in the future.

Observations of the time delay are given in Table \ref{tab3}. By comparing the results with those in Ref.\cite{virbhadra}, we can see that both the relativistic Einstein ring (when $\beta=0$) and the time delay become larger with the phantom constant $b$.
\begin{table}[h]
\caption{Numerical values for the offset of the deflection angle $\Delta\alpha_{1p}$ and time delay $\tau_{1p}$ of the first relativistic images (on the same side as the primary image) with different values of constant $b$ and angular source position $\beta$. Here $\beta$, $\Delta\alpha_{1p}$  and $\tau_{1p}$ are, respectively, expressed in arcseconds (arcsec), microarcseconds ($\mu$as), and minutes (min). In order to compare the results with those in Ref.\cite{virbhadra}, we here apply the most recent data for our Galaxy \cite{eisenhauer}. We suppose that the mass $M=3.61\times10^6M_\odot$ and the distance $D_{\text{OL}}=7.62$ kpc, so that $M/D_{OL}\approx2.26\times10^{-11}$ and $D_{\text{OL}}/D_{\text{OS}}=1/2$.}\label{tab3}
\begin{center}
\begin{tabular}{ccccccccccc}
\hline \hline  &\multicolumn{2}{c}{$b=0.02$}&\multicolumn{2}{c}
{$b=0.5$}
&\multicolumn{2}{c} {$b=1.0$}&\multicolumn{2}{c} {$b=2.0$}&\multicolumn{2}{c} {$b=3\pi/4$} \\
 $\beta$(arcsec)&$\Delta\alpha_{1p}(\mu$as)&$\tau_{1p}$(min)
 &$\Delta\alpha_{1p}(\mu$as)&$\tau_{1p}$(min)
 &$\Delta\alpha_{1p}(\mu$as)&$\tau_{1p}$(min)&$\Delta\alpha_{1p}(\mu$as)&$\tau_{1p}$(min)
 &$\Delta\alpha_{1p}(\mu$as)&$\tau_{1p}$(min)
  \\ \hline
$0$       &48.60849&38.3701&50.17959&38.4742&54.45789&39.3316&67.78770&42.3771&73.35538&43.5699 \\
$10^{-6}$ &46.60849&38.3702&48.17959&38.4742&52.45789&39.3316&65.78770&42.3771&71.35538&43.5699\\
$10^{-5}$ &28.60848&38.3703&30.17959&38.4742&34.45789&39.3316&47.78770&42.3771&53.35538&43.5699 \\
$10^{-4}$
&$-151.392$&38.3704&$-149.820$&38.4742&$-145.542$&$39.3316$&$-132.212$&42.3771&$-126.645$
&43.5699\\
$10^{-3}$
&$-1951.39$&38.3706&$-1949.82$&38.4742&$-1945.54$&39.5665&$-1932.21$&42.3771&$-1926.64$
&43.5699\\
$10^{-2}$
          &$-19951.4$&$38.3708$&$-19949.8$&$38.4742$&$-19945.5$&$39.5695$
&$-19932.2$&$42.3772$&$-19926.6$&$43.5699$\\
 $10^{-1}$&$-199951$&$38.3732$&$-199949$&$38.4772$&$-199945$&$39.5752$
&$-199932$&$42.3802$&$-199926$&$43.5730$\\
$1$
         &$-1999951$&$38.4752$&$-1999949$&$38.7812$&$-1999945$&$39.8736$
&$-1999932$&$42.6841$&$-1999926$&$43.8769$\\
\hline\hline
\end{tabular}
\end{center}
\end{table}
\section{Summary}
Modern observations show that the Universe is expanding with an acceleration that is dominated by a peculiar kind of matter (e.g., dark energy) which can be modeled by quintessence or phantom scalar fields. This unknown matter has unusual properties such as negative values of the pressure-to-density ratio. If it exists, then it will inevitably affect the known spacetimes such as black hole physics.  Studying strong gravitational lensing can help us probe its existence and properties from astronomical observations. We have investigated strong-field lensing in the phantom black hole spacetime to study the influence of the phantom constant on strong gravitational lensing. The model was applied to the supermassive black hole at the Galactic Center.

Our results show that with an increase of the phantom constant $b$ both the minimum
impact parameter $u_{\text{ps}}$ and the angular position of the relativistic
images $\theta_{\infty}$ increase, which contrary to the case of Reissner-Norstr\"{o}m black hole lensing with an electric charge $q$. The photon sphere $r_{\text{ps}}$ and relative magnitudes $r_m$ stay the same as those of a Schwarzschild  black hole and are independent on $b$, which is also contrary to the case where they are weakened by an electric charge. However, the deflect angle $\alpha(\theta)$ and the angular separation $s$ appear to have similar behavior as  in Reissner-Norstr\"{o}m lensing.   This may offer a way to distinguish a phantom black hole from a Reissner-Norstr\"{o}m one using the astronomical instruments developed in the future.

Our results also show the effects of dark energy in the considered model, such as (i)  compressing the usual black hole and more powerfully attracting photons, (ii) making the relativistic Einstein ring  larger than the usual black hole, and (iii) not weakening the usual  relative magnitudes, which will facilitate observations.

\vspace{0.5cm}
Two days after this paper was published to the arXiv, Ref. \cite{eiroa} appeared online in the same database, containing a partial overlap with our work.

\begin{acknowledgments}
This work was supported by the National Natural Science Foundation of China
under No. 11247013, the Hunan Provincial NSFC No. 11JJ3014, the Scientific
Research Fund of the Hunan Provincial Education Department No. 11B067, the
Foundation for the Author of Hunan Provincial Excellent Doctoral Dissertation
No. YB2012B034, and the Aid program for Science and Technology Innovative Research
Team in Higher Educational Institutions of Hunan Province.
\end{acknowledgments}

\appendix
\begin{center}{\bf Appendix: Energy density and pressure of phantom fields}\end{center}

With the metric (\ref{metric}), the components for the energy-momentum tensor of phantom fields are
\begin{eqnarray}
T_0^0&=&-\frac{rf'}{b^2+r^2}-\frac{(2b^2+r^2)f}{(b^2+r^2)^2}+\frac{1}{b^2+r^2},\\ \nonumber
T_1^1&=&-\frac{rf'}{b^2+r^2}-\frac{r^2f}{(b^2+r^2)^2}+\frac{1}{b^2+r^2},\\ \nonumber
T_2^2&=&T_3^3=-\frac{rf'}{b^2+r^2}-\frac{b^2f}{(b^2+r^2)^2}-\frac{f''}{2}.
\end{eqnarray}
We can rewrite them as a appropriate general expression \cite{kiselev},
\begin{eqnarray}
T_0^0&=&\rho_p(r),\;\;\\ \nonumber
T_i^j&=&C(r)r_ir^j+B(r)\delta_i^j\\ \nonumber
&=&3\rho_p(r)\omega_p\Big[-(1+3D)\frac{r_ir^j}{r_nr^n}+D\delta_i^j\Big],
\end{eqnarray}
so that the spatial part is proportional to the time component with the arbitrary parameter $D$
depending on the internal structure of phantom fields. An isotropic averaging over the angles
gives
\begin{eqnarray}
\langle T_i^j\rangle&=&-\rho_p(r)\omega_p\delta_i^j=-p_p(r)\delta_i^j,
\end{eqnarray}
and therefore $p_p(r)=\omega_p\rho_p(r)$. After such a treatment, the results are
\begin{eqnarray}
T_1^1&=&\rho_p(r)+\frac{2b^2f}{(b^2+r^2)^2},\;\;\\ \nonumber
T_2^2&=&T_3^3=-\frac{1}{2}(3\,\omega_p+1)\rho_p(r)-\frac{b^2f}{(b^2+r^2)^2}.
\end{eqnarray}
At last, we obtain the expressions for the pressure and pressure-to-energy density ratio
\begin{eqnarray}
\omega_p&=&\frac{(2b^2+r^2)f+(b^2+r^2)\big[(b^2+r^2)f''+3rf'-1\big]}
{3\big[(2b^2+r^2)f+(b^2+r^2)(rf'-1)\big]},\;\;\\ \nonumber
p_p(r)&=&\frac{rf'}{b^2+r^2}+\frac{f''}{3}+\frac{(2b^2+r^2)f}{3(b^2+r^2)^2}-\frac{1}{3(b^2+r^2)}.
\end{eqnarray}

\end{document}